\documentclass
[superscriptaddress,secnumarabic,amssymb,amsmath,nobibnotes,aps,prd,showkeys,nofootinbib,onecolumn,notitlepage]{revtex4}%
\usepackage{setspace}
\usepackage{color}
\usepackage{amsmath}
\usepackage{amsfonts}
\usepackage{verbatim}
\usepackage{amssymb}
\usepackage{graphicx,bm}
\usepackage{graphicx}
\usepackage{amsmath}
\usepackage{amssymb}
\usepackage{amssymb}
\usepackage{graphicx,bm}
\usepackage{graphicx}
\usepackage[caption=false]{subfig}%
\providecommand{\U}[1]{\protect\rule{.1in}{.1in}}
\newcommand{\be}{\begin{equation}}
\newcommand{\ee}{\end{equation}}

\newcommand{\mincir}{\raise
-3.truept\hbox{\rlap{\hbox{$\sim$}}\raise4.truept\hbox{$<$}\ }}
\newcommand{\magcir}{\raise
-3.truept\hbox{\rlap{\hbox{$\sim$}}\raise4.truept\hbox{$>$}\ }}

\ifx\pdfoutput\relax\let\pdfoutput=\undefined\fi
\newcount\msipdfoutput
\ifx\pdfoutput\undefined\else
\ifcase\pdfoutput\else
\ifx\paperwidth\undefined\else
\ifdim\paperheight=0pt\relax\else\pdfpageheight\paperheight\fi
\ifdim\paperwidth=0pt\relax\else\pdfpagewidth\paperwidth\fi
\fi\fi\fi
\begin{document}
\title{Quantum potentiality in Inhomogeneous Cosmology}
\author{Andronikos Paliathanasis}
\email{anpaliat@phys.uoa.gr}
\affiliation{Institute of Systems Science, Durban University of Technology, Durban 4000,
South Africa}

\begin{abstract}
For the Szekeres system which describes inhomogeneous and anisotropic
spacetimes we make use of a point-like Lagrangian, which describes the
evolution of the physical variables of the Szekeres model, in order to perform
a canonical quantization and to study the quantum potentiality of the Szekeres
system in the content of de Broglie--Bohm theory. We revise previous results
on the subject and we find that for a specific family of trajectories with
initial conditions which satisfy a constraint equation, there exists
additional conservation laws for the classical Szekeres system which are used
to define differential operators and to solve the Wheeler-DeWitt equation.
From the new conservation laws we construct a wave function which provides a
nonzero quantum potential term that modifies the Szekeres system. The quantum
potential corresponds to new terms in the dynamical system such that new
asymptotic solutions with a nonzero energy momentum tensor of an anisotropic
fluid exist. Therefore, the silent property of the Szekeres spacetimes is
violated by quantum correction terms.

\end{abstract}
\keywords{Bohmian mechanics, quantum cosmology, inhomogeneous spacetimes, exact
solutions, Szekeres universes}\maketitle
\date{\today}

\section{Introduction}

\label{sec1}

Quantum gravity is motivated by the idea to have a quantum description of all
the matter fields and of their interactions in a gravitational system
\cite{rqc1}. The dynamical variables of the spacetime interact with the matter
fields as is described by General Relativity.\ Hence, in quantum gravity the
dynamical variables of the spacetimes are described by quantum physics. The
study of the quantum properties of the whole universe as a unique
gravitational system is part of quantum cosmology. For extended discussions we
refer the reader to \cite{rqc1,rqc2}. In this work we are interested in the
analytic solutions of the Wheeler-DeWitt equation \cite{wdw1,wdw2} for quantum
cosmology in the case of inhomogeneous spacetimes and in the derivation of
quantum corrections on the semiclassical limit as described by de
Broglie--Bohm theory \cite{bm1,bm2,bm3,bm4,bm5}.

Inhomogeneous cosmological models are exact spacetimes which in general do not
admit any isometry vector field while the conditions described by the
cosmological principle, that is, the limit of the
Friedmann--Lema\^{\i}tre--Robertson--Walker universe is provided \cite{kras}.
Inhomogeneous cosmological models can be used for the description of the
universe in the preinflationary era \cite{dd1,dd2,dd3,dd4,dd5}, as also in the
description of the small inhomogeneities which are found by cosmological
observations \cite{dd6,dd7,dd8,dd9,dd10}.\ This is an alternative approach
based on exact solutions and is different from the cosmological perturbation theory.

In the following we are interested in Szekeres spacetimes \cite{szek}. These
spacetimes are exact solutions of the gravitational field equations for a
diagonal line element with two dynamical variables and zero magnetic component
for the Weyl tensor. The matter field is described by an inhomogeneous
pressureless fluid. Szekeres spacetimes are characterized as \textquotedblleft
partially locally rotational\textquotedblright\ spacetimes \cite{silent0}. The
term \textquotedblleft partially\textquotedblright\ indicates that the exact
solutions do not admit any isometry vector field. Furthermore, there is no
information dissemination with gravitational or sound waves on these exact
solutions which means that they are silent universes \cite{silent}. There are
various interesting results in the literature which have shown that Szekeres
spacetimes can play a significant role in the description of various epochs of
our universe \cite{per101,per102,per103,per104}, while the present isotropic
and homogeneous on large scales observed universe can be provided by Szekeres
spacetimes for specific initial conditions with or without an inflationary era
in the cosmological evolution \cite{bbb01,bbb02,bbb03}. Various
generalizations of the Szekeres spacetimes with other kind of matter fields
can be found in \cite{sz1,sz2,sz3,sz4,sz9,sz10,sz11}.

Szekeres universes are described by\ (pseudo)-Riemannian geometry with line
element $ds^{2}=g_{\mu\nu}dx^{\mu}dx^{\nu}$,~$x^{\mu}=\left(  t,x,y,z\right)
$ and%
\begin{equation}
g_{\mu\nu}=diag\left(  -1,e^{2\alpha\left(  t,x,y,z\right)  },e^{2\beta\left(
t,x,y,z\right)  },e^{2\beta\left(  t,x,y,z\right)  }\right)  ,
\end{equation}
where functions $\alpha\left(  t,x,y,z\right)  $ and $\beta\left(
t,x,y,z\right)  $ satisfy the following algebraic-differential system known as
the Szekeres system.\ The first-order differential equations are%
\begin{align}
\dot{\rho}+\theta\rho &  =0,\label{sz.01}\\
\dot{\theta}+\frac{\theta^{2}}{3}+6\sigma^{2}+\frac{1}{2}\rho &
=0,~\label{sz.02}\\
\dot{\sigma}-\sigma^{2}+\frac{2}{3}\theta\sigma+E  &  =0,\label{sz.03}\\
\dot{E}+3E\sigma+\theta E+\frac{1}{2}\rho\sigma &  =0\label{sz.04}%
\end{align}
with the algebraic constraint%

\begin{equation}
\frac{\theta^{2}}{3}-3\sigma^{2}+\frac{^{\left(  3\right)  }R}{2}%
=\rho,\label{sz.05}%
\end{equation}
{where $\theta=\left(  \frac{\partial\alpha}{\partial t}\right)  +2\left(
\frac{\partial\beta}{\partial t}\right)  ~~$ and $~\sigma^{2}=\frac{2}%
{3}\left(  \left(  \frac{\partial\alpha}{\partial t}\right)  -\left(
\frac{\partial\beta}{\partial t}\right)  \right)  ^{2}$~are the expansion rate
and the anisotropic parameter, }$\rho=\rho\left(  t,x,y,z\right)  $ is the
energy density of the inhomogeneous dust fluid and $E=E\left(  t,x,y,z\right)
$ is the electric component of the Weyl tensor~{$E_{\nu}^{\mu}=$ $Ee_{\nu
}^{\mu};$} and $^{\left(  3\right)  }R$ is the spatial curvature of the
three-dimensional hypersurface. In addition to the latter system, the
dynamical variables satisfy the propagation equations which are {$h_{\mu}%
^{\nu}\sigma_{\nu;\alpha}^{\alpha}=\frac{2}{3}h_{\mu}^{\nu}\theta_{;\nu}%
$~,$~h_{\mu}^{\nu}E_{\nu;\alpha}^{\alpha}=\frac{1}{3}h_{\mu}^{\nu}\rho_{;\nu}$
in which $h_{\mu\nu}$ is the decomposable tensor defined by the expression
$h_{\mu\nu}=g_{\mu\nu}-u_{\mu}u_{\nu}$, where }$u^{\mu}$ is a unitary vector
field, $u^{\mu}u_{\mu}=-1,$ which defines the physical observer \cite{lesame}.

The integrability properties of the Szekeres system have been widely studied
in the literature \cite{isz1,isz2,isz3}, In \cite{isz1} the Szekeres system
has been written as a system of two second-order differential equations and a
conservation law quadratic in the derivatives of the dynamical variables was
derived. The gravitational field equations of the Szekeres model do not admit
a minisuperspace description. However, as was shown in \cite{isz1}, the
time-dependent field equations can be described by a point-like Lagrangian
with respect to the variables $\left\{  \rho,E\right\}  $, while the
conservation law quadratic in the derivatives is derived easily with the use
of Noether's theorem. In addition in this work we show that the Szekeres
system admits as an additional conservation law the Lewis invariant.

The Lagrangian description of the Szekeres system and the Noetherian
conservation law were applied in \cite{isz4} to quantize and write the
\textquotedblleft time\textquotedblright-independent Wheeler-DeWitt equation
for the Szekeres system. The solution of the Wheeler-DeWitt equation which
satisfies the quadratic conservation law has been found which does not affect
the classical trajectories of the Szekeres system and there is no any quantum
potential term provided. Moreover, a probability function was found from which
it was found that a stationary surface of the probability function is related
with classical exact solutions. A similar study was performed recently for the
Szekeres system with the cosmological constant term \cite{isz5}.

In the following we revise the analysis of \cite{isz4}. Specifically, we find
that, in the special case for which the \textquotedblleft
energy\textquotedblright\ of the two-dimensional Hamiltonian system which
describes the Szekeres system vanishes, new conservation laws exist for the
Szekeres system. These new conservation laws are used to define \ quantum
operators which are used as supplementary conditions on the Wheeler-DeWitt
equation and to determine new similarity solutions for the Wheeler-DeWitt
equation \cite{qu3,qu4,qu5}. From the new wavefunction we are able to
construct a nontrivial quantum potential given by the Broglie--Bohm theory.
The effects of the quantum correction in the original Szekeres system is
investigated as also the effects on the dynamics are studied. The plan of the
paper is as follows.

In Section \ref{sec2}, we present the two-dimensional Hamiltonian dynamical
system which is equivalent to the Szekeres system (\ref{sz.01})-(\ref{sz.04})
and we derive the new conservation laws. In particular, we show that the
Szekeres system admits as conservation law the Lewis invariant as also a
family of conservation laws generated by Lie point symmetries when the energy
of the Hamiltonian dynamical system is zero. The new conservation laws are
applied for the derivation of quantum operators and for the derivation of
exact solutions of the Wheeler-DeWitt equation in Section \ref{sec3}. The
quantum potential is determined in \ref{sec4}, in which we show that a nonzero
quantum correction exists for specific trajectories of the Szekeres system
satisfying a specific set of initial conditions. In Section \ref{sec5}, we
study the effects of the nonzero quantum correction in the Szekeres system. We
write the modified Szekeres system for which a contribution in the equation
for the expansion rate $\theta~$follows by quantum corrections. This new term
modifies the dynamics of the Szekeres system and leads to new asymptotic
solutions with nonzero matter, pressure and anisotropic component of an energy
momentum tensor. These latter fluid components have their origin in the
quantum corrections as provided by Bohmian mechanics. Finally, in Section
\ref{sec6}, we summarize our results and we draw our conclusions.

\section{Hamiltonian formulation of the Szekeres system}

\label{sec2}

From equations (\ref{sz.01}) and (\ref{sz.04}) we find that%
\begin{equation}
\theta=-\frac{\dot{\rho}}{\rho}~,~\sigma=\frac{2\left(  \dot{\rho}%
E\mathcal{-}\rho\dot{E}\right)  }{\rho\left(  \rho+6E\right)  }~.\label{sz.06}%
\end{equation}
Thus by substituting (\ref{sz.06}) into (\ref{sz.02}) and (\ref{sz.03}) we
obtain an equivalent form for the Szekeres system comprising two second-order
differential equations with respect to the variables $\rho$ and $E$. The
resulting equations are \cite{isz1}%
\begin{align}
\ddot{u}+\frac{1}{u^{2}}  &  =0,\label{sz.07}\\
\ddot{v}-2\frac{v}{u^{3}}  &  =0,\label{sz.08}%
\end{align}
where $\left\{  u,v\right\}  $ are new variables defined as \cite{isz1}%
\begin{equation}
\rho\left(  u,v\right)  =\frac{6}{\left(  u-v\right)  u^{2}}~,\ E\left(
u,v\right)  =\frac{v}{u^{3}\left(  v-u\right)  }\label{sz.09}%
\end{equation}
with inverse transformation%
\begin{equation}
u\left(  \rho,E\right)  =\left(  \frac{\rho}{6}+E\right)  ^{-\frac{1}{3}%
}~,~v\left(  \rho,E\right)  =-\frac{6E}{\rho}\left(  \frac{\rho}{6}+E\right)
^{-\frac{1}{3}}.\label{sz.10}%
\end{equation}
The parameters $\theta$ and $\sigma$, are expressed in terms of the new
variables as
\begin{align}
\theta\left(  u,v,\dot{u},\dot{v}\right)   &  =\frac{\left(  3u-2v\right)
\dot{u}-u\dot{v}}{u\left(  u-v\right)  },\label{sz.10a}\\
\sigma\left(  u,v,\dot{u},\dot{v}\right)   &  =\frac{u\dot{v}-v\dot{u}}%
{3u^{2}-3uv}~.\label{sz.10b}%
\end{align}

Equation (\ref{sz.07}) can be integrated by quadratures, that is, $\int
\frac{du}{\sqrt{2u^{-1}+I_{0}}}=t-t_{0}$, where $I_{0}$ is a constant of
integration and a conservation law for the dynamical system (\ref{sz.07}%
)-(\ref{sz.08}). That is, $I_{0}=\dot{u}^{2}-2u^{-1}$. Thus by writing
$u=u\left(  t\right)  $ in (\ref{sz.08}) we obtain a linear equation of the
form $\ddot{v}+\omega\left(  t\right)  v=0$, $\omega\left(  t\right)
=-2u\left(  t\right)  ^{-3}$. \ This linear second-order equation is known as
the time-dependent oscillator \cite{lt1} and it is a maximally symmetric
equation. Moreover, it admits as conservation law the Lewis invariant given by
the expression \cite{lt2,lt3,lt4}
\begin{equation}
\Phi=\frac{1}{2}\left(  \left(  y\dot{v}-\dot{y}v\right)  ^{2}+\left(
\frac{v}{y}\right)  ^{2}\right)  ,\label{sz.11}%
\end{equation}
where $y\left(  t\right)  $ is any solution of the Ermakov-Pinney differential
equation~\cite{lt5}, that is,%
\begin{equation}
\ddot{y}+\omega\left(  t\right)  y=\frac{1}{y^{3}}.\label{sz.12}%
\end{equation}

Conservation law (\ref{sz.11}) is a new conservation law found for the
Szekeres system and has not been derived before. It is really a point of
interest that the Lewis invariant and the Ermakov-Pinney equation appears in
inhomogeneous cosmology. Previously, the Ermakov-Pinney equation has appeared
and in other gravitational such in scalar tensor theories and in modified
theories of gravity \cite{erm1,erm2}.

It is easy to observe that the Szekeres equations (\ref{sz.07}), (\ref{sz.08})
can be derived by the variation of the point-like Lagrangian \cite{isz1}%
\begin{equation}
L\left(  u,\dot{u},v,\dot{v}\right)  =\dot{u}\dot{v}-\frac{v}{u^{2}%
}\label{sz.13}%
\end{equation}
with Hamiltonian function
\begin{equation}
\mathcal{H}=p_{u}p_{v}+\frac{v}{u^{2}}\equiv h\left(  u,v,p_{u},p_{v}\right) ,
~\label{sz.14}%
\end{equation}
where
\begin{equation}
\dot{u}=p_{v}~,~\dot{v}=p_{u}~\label{sz.15}%
\end{equation}
and
\begin{equation}
\dot{p}_{v}=-\frac{1}{u^{2}}~,~\dot{p}_{v}=2\frac{v}{u^{3}}.\label{sz.16}%
\end{equation}
That is a Hamiltonian description for the Szekeres system, for which we can
see that (\ref{sz.14}) is a conservation law with $h\left(  u,v,p_{u}%
,p_{v}\right)  =const.$, which corresponds to the \textquotedblleft
energy\textquotedblright\ of the Hamiltonian system, because equations
(\ref{sz.07}), (\ref{sz.08}) are autonomous. In terms of the momentum the
conservation law $I_{0}$ becomes $I_{0}=p_{v}^{2}-2u^{-1}$, which is in
involution and independent of the Hamiltonian $h$. Hence, as has found before
the Szekeres system is Liouville integrable \cite{isz1}. The conservation law
$I_{0}$ is related with a generalized symmetry which generates a constant
transformation for the dynamical system in which the Action Integral for the
Szekeres system remains invariant, that is, $I_{0}$ is a Noetherian
conservation law. That is not true for the Lewis invariant (\ref{sz.11}).

The application of Noether's theorem for point and contact transformations of
the Hamiltonian system (\ref{sz.14}), (\ref{sz.15}) and (\ref{sz.16}) as was
found in \cite{isz1} does not provide additional conservation laws. However,
the existence of conservation laws for the trajectories with initial
conditions $h=0$ has been investigated before. When $h=0$, the trajectories of
the Szekeres system can be seen as \textquotedblleft
null-like\textquotedblright\ trajectories which are conformally invariant, see
the discussion in \cite{mtnull,nnull,nnull2}.

Therefore, when $h\equiv0$, we find that the Szekeres system (\ref{sz.07}),
(\ref{sz.08}) admits the additional conservation laws%

\begin{equation}
I_{1}=u^{2}\dot{v}~\text{or }I_{1}=u^{2}p_{u}~,\label{sz.17}%
\end{equation}%
\begin{equation}
I_{2}=\frac{\dot{u}}{v}~\text{or }I_{2}=\frac{p_{v}}{v}~,\label{sz.18}%
\end{equation}
and%
\begin{equation}
I_{3}=2u\dot{v}+v\dot{u}~\text{or~}I_{3}=2up_{u}+vp_{v}.\label{sz.19}%
\end{equation}
These conservation laws can be derived by the application of Noether's theorem
for a conformally related Lagrangian of function (\ref{sz.13}), while they are
generated by conformal transformations of the two-dimensional flat space which
defines the kinetic energy for the Lagrangian function (\ref{sz.13}).

Consider now the conformal transformation $dt=u^{2}d\tau$. The conformally
equivalent Lagrangian of (\ref{sz.13}) is given by
\begin{equation}
\bar{L}\left(  u,\frac{du}{d\tau},v,\frac{d\dot{u}}{d\tau}\right)  =\frac
{1}{u^{2}}\left(  \frac{du}{d\tau}\right)  \left(  \frac{dv}{d\tau}\right)
-v,\label{sz.20}%
\end{equation}
where now the Szekeres system becomes%
\begin{equation}
\ddot{u}-\frac{2}{u}\dot{u}^{2}+u^{2}=0~,~\ddot{v}=0.\label{sz.21}%
\end{equation}
Under the conformal transformation the conservation law $\bar{I}_{1}$ becomes
$\bar{I}_{1}=\dot{v}$. Hence, with the use of the constraint equation
$\frac{1}{u^{2}}\left(  \frac{du}{d\tau}\right)  \left(  \frac{dv}{d\tau
}\right)  +v=0$, the closed-form solution of the Szekeres system is expressed
as
\begin{equation}
u\left(  \tau\right)  =2\left(  \left(  \tau-\tau_{0}\right)  ^{2}%
+u_{1}\right)  ^{-1}~,~v\left(  t\right)  =\bar{I}_{1}\left(  \tau-\tau
_{0}\right)  \text{.}\label{sz.22}%
\end{equation}

Hence, we can write the closed-form solution for the original variables%
\begin{equation}
\rho\left(  \tau\right)  =\frac{3\left(  \left(  \tau-\tau_{0}\right)
^{2}+u_{1}\right)  ^{3}}{2\left(  I_{1}\left(  \tau-\tau_{0}\right)  \left(
\left(  \tau-\tau_{0}\right)  ^{2}+u_{1}\right)  -2\right)  }~,\label{sz.23}%
\end{equation}%
\begin{equation}
E\left(  \tau\right)  =\frac{I_{1}\left(  \tau-\tau_{0}\right)  \left(
\left(  \tau-\tau_{0}\right)  ^{2}+u_{1}\right)  ^{4}}{8\left(  I_{1}\left(
\tau-\tau_{0}\right)  \left(  \left(  \tau-\tau_{0}\right)  ^{2}+u_{1}\right)
-2\right)  }~,\label{sz.24}%
\end{equation}%
\begin{equation}
\theta\left(  \tau\right)  =\frac{\left(  \left(  \tau-\tau_{0}\right)
^{2}+u_{1}\right)  \left(  I_{1}\left(  \left(  \tau-\tau_{0}\right)  \left(
\left(  \tau-\tau_{0}\right)  ^{2}+u_{1}\right)  ^{3}-4\right)  -12\left(
\tau-\tau_{0}\right)  \right)  }{8-I_{1}\left(  \left(  \tau-\tau_{0}\right)
^{2}+u_{1}\right)  ^{3}}~,\label{sz.25}%
\end{equation}%
\begin{equation}
\sigma\left(  \tau\right)  =\frac{I_{1}\left(  \left(  \tau-\tau_{0}\right)
^{2}+u_{1}\right)  \left(  8+\left(  \tau-\tau_{0}\right)  \left(  \left(
\tau-\tau_{0}\right)  ^{2}+u_{1}\right)  \right)  }{6\left(  8-I_{1}\left(
\left(  \tau-\tau_{0}\right)  ^{2}+u_{1}\right)  ^{3}\right)  }.\label{sz.26}%
\end{equation}

In a similar way closed-form solutions can be found for other conformal time
$dt\rightarrow d\bar{\tau}$, by using the remaining vector fields. However,
they are the same solutions expressed in different coordinates. Moreover, it
is important to mention here that the constants of integration and the
conservation laws, are constants with respect to the derivative in time, that
is, they are functions of the original coordinates of the spacetime, for
instance $t_{0}=t_{0}\left(  x,y,z\right)  $ and $t_{0}$ is an essential
integration parameter that is why we have not omitted it.

We continue our analysis by performing a canonical quantization for the
Hamiltonian (\ref{sz.14}) to write the Wheeler-DeWitt equation for the
Szekeres system, while we use the conservation laws related with point and
contact symmetries to solve the Wheeler-DeWitt equation and write the wave function.

\section{The Wheeler-DeWitt equation}

\label{sec3}

In terms of the $3+1$ decomposition notation of General Relativity, the
Wheeler-DeWitt equation follows from the Hamiltonian constraint of the field
equations \cite{wdw1}. The Wheeler-DeWitt equation is not a single
differential equation, but it defines a family of equations where at every
point of the 3-dimensional hypersurfaces a unique equation is defined.
However, in the case of the minisuperspace approximation the infinite degrees
of freedom of the superspace reduce to a finite number. Hence, instead of
having an equation for each point of the hypersurface, there follows a unique
equation for all of the points \cite{wdw2}.

The Szekeres system (\ref{sz.01})-(\ref{sz.04}) does not admit a
minisuperspace description. However, through the dynamical variables it can be
written as a two-dimensional Hamiltonian system with constraint (\ref{sz.14}).
By using that property we are able to study the quantization of the Szekeres
system. \ Specifically we perform a canonical quantization by promoting the
Poisson brackets to commutators and the variables on the phase space into
operators $x^{i}\rightarrow\hat{x}^{i}=x^{i},\ p_{i}\rightarrow\hat{p}%
_{i}=i\frac{\partial}{\partial x^{i}}$. Thus from the Hamiltonian
(\ref{sz.14}) there follows the time-independent Schr\"{o}dinger equation
\cite{isz4}%
\begin{equation}
\left(  -\frac{\partial}{\partial u\partial v}+\frac{v}{u^{2}}-h\right)
\Psi\left(  u,v\right)  =0.\label{sz.27}%
\end{equation}

At this point we can use the conservation laws to define operators which keep
invariant the Wheeler-DeWitt equation (\ref{sz.27}). For arbitrary value of
$h$ from $I_{0}$, there follows the quantum operator%
\begin{equation}
\left(  \frac{\partial}{\partial v^{2}}+\frac{2}{u}+I_{0}\right)  \Psi\left(
u,v\right)  =0~,\label{sz.28}%
\end{equation}
while, when $h=0$, the additional operators%
\begin{equation}
\left(  u^{2}\frac{\partial}{\partial u}+iI_{1}\right)  \Psi\left(
u,v\right)  =0~,\label{sz.29}%
\end{equation}%
\begin{equation}
\left(  \frac{1}{v}\frac{\partial}{\partial v}+iI_{2}\right)  \Psi\left(
u,v\right)  =0~,\label{sz.30}%
\end{equation}%
\begin{equation}
\left(  2u\frac{\partial}{\partial u}+v\frac{\partial}{\partial v}%
+iI_{3}\right)  \Psi\left(  u,v\right)  =0\label{sz.31}%
\end{equation}
exist.

With the use of one of these differential operators we can construct a
solution for the Wheeler-DeWitt equation (\ref{sz.27}). These solutions are
called similarity solutions because the differential operators are related
with Lie symmetries for the differential equation (\ref{sz.27}).

In \cite{isz4} the differential operator (\ref{sz.28}) was applied, which
provides the wavefunction%
\begin{equation}
\Psi_{A}\left(  h,I_{0},u,v\right)  =\frac{\sqrt{u}}{\sqrt{2+I_{0}u}}\left(
\Psi_{A}^{1}\cos f\left(  u,v\right)  +\Psi_{A}^{2}\sin f\left(  u,v\right)
\right)  ,\label{sz.32}%
\end{equation}
where~~$f\left(  u,v\right)  =\left(  I_{0}^{3/2}\sqrt{u}\right)  ^{-1}\left(
(hu+I_{0}v)\sqrt{2I_{0}+I_{0}^{2}u}-2h\sqrt{u}\text{\textrm{arcsin}}%
\mathrm{h}\left(  \sqrt{\frac{I_{0}{u}}{2}}\right)  \right)  ~,~$for
$I_{0}\neq0,$ or$~f\left(  u,v\right)  =\frac{\sqrt{2}}{3}\left(
hu^{2}+3v\right)  u^{-\frac{1}{2}},$ for $I_{0}=0.$ Coefficients $\Psi
_{A}^{1,2}$ are constants of integration. In a similar way we can construct
similarity solutions by using the other differential operators. For more
details on the properties of the wavefunction (\ref{sz.32}) we refer the
reader in \cite{isz4}.

For $h=0$, the wave function, $\Psi\left(  u,v\right) , $ which satisfies the
differential operator (\ref{sz.29}) is%
\begin{equation}
\Psi_{B}\left(  I_{1},u,v\right)  =\Psi_{B}^{1}\exp\left(  -\frac{i}{2I_{1}%
}\left(  v^{2}+2\frac{\left(  I_{1}\right)  ^{2}}{u}\right)  \right)
,\label{sz.33}%
\end{equation}
while from (\ref{sz.30}) we find that
\begin{equation}
\Psi_{C}\left(  I_{2},u,v\right)  =\Psi_{C}^{1}\exp\left(  \frac{i}{2I_{2}%
}\left(  \frac{2}{u}+\left(  I_{2}\right)  ^{2}v^{2}\right)  \right)
.\label{sz.34}%
\end{equation}
Furthermore from (\ref{sz.31}) there follows the wave function%
\begin{equation}
\Psi_{D}\left(  \beta,u,v\right)  =\left(  v^{2}u\right)  ^{\frac{\beta}{4}%
}\left(  \Psi_{D}^{1}J_{\frac{\beta}{2}}\left(  v\sqrt{\frac{2}{u}}\right)
+\Psi_{D}^{2}Y_{\frac{\beta}{2}}\left(  v\sqrt{\frac{2}{u}}\right)  \right)
\label{sz.35}%
\end{equation}
in which $J_{\alpha}\left(  u,v\right)  $ and $Y_{\alpha}\left(  u,v\right)  $
are the Bessel functions of the first and of the second kinds, respectively,
and $I_{3}=-i\beta$.

In the limit for which $\sqrt{\frac{2}{u}}v\rightarrow\infty\,$, the wave
function (\ref{sz.35}) is approximated by the functional form%
\begin{equation}
\hat{\Psi}_{D}\left(  \beta,u,v\right)  =v^{\frac{\beta-1}{2}}u^{\frac
{\beta+1}{4}}\left(  \hat{\Psi}_{D}^{1}\exp\left(  i\left(  \sqrt{\frac{2}{u}%
}v-\left(  \beta+1\right)  \frac{\pi}{4}\right)  \right)  +\hat{\Psi}_{D}%
^{2}\exp\left(  -i\left(  \sqrt{\frac{2}{u}}v-\left(  \beta+1\right)
\frac{\pi}{4}\right)  \right)  \right)  ,\label{sz.36}%
\end{equation}
where $\hat{\Psi}_{D}^{1},~\hat{\Psi}_{D}^{2}$ are constants. Function
$\hat{\Psi}_{D}\left(  \beta,u,v\right)  $ describes oscillations of the polar
form $\Psi\left(  u,v\right)  =\Omega\left(  u,v\right)  e^{iS\left(
u,v\right)  }$ with amplitude~$\Omega\left(  u,v\right)  =v^{\frac{\beta-1}%
{2}}u^{\frac{\beta+1}{4}}$ and radial argument $S\left(  u,v\right)  =\left(
\sqrt{\frac{2}{u}}v-\left(  \beta+1\right)  \frac{\pi}{4}\right)  $.

\section{Quantum potential}

\label{sec4}

In Bohmian quantum theory, the main difference from the classical theory is
the quantum Hamilton-Jacobi equation which for a wave function expressed in
the Madelung representaiton $\Psi\left(  \mathbf{y}\right)  =\Omega\left(
\mathbf{y}\right)  e^{iS\left(  \mathbf{y}\right)  }$ is defined as%
\begin{equation}
\frac{1}{2}G^{AB}\partial_{A}S\left(  \mathbf{y}\right)  \partial_{B}S\left(
\mathbf{y}\right)  +V\left(  \mathbf{y}\right)  +h+Q_{V}\left(  \mathbf{y}%
\right)  =0,\label{sz.37}%
\end{equation}
where the term $Q_{V}\left(  \mathbf{y}\right)  ~$ is known as the quantum
potential and is related with the amplitude of the wave function
\begin{equation}
Q_{V}\left(  \mathbf{y}\right)  =-\frac{\Box\Omega\left(  \mathbf{y}\right)
}{2\Omega\left(  \mathbf{y}\right)  }\label{sz.38}%
\end{equation}
and $\Box~$is the Laplacian of the time-independent Schr\"{o}dinger equation,
that is, for our model, $\Box=\frac{\partial}{\partial u\partial v}$.{\ The
radial argument play the role of the Action, so that the canonical momentum is
given as }$p_{A}=\frac{\partial S}{\partial y^{A}}$. In the WKB approximation
the quantum potential it is neglected and the classical limit is recovered.
The Madelung representation is an alternative way to write the Schr\"{o}dinger
equation in a real and complex imaginary part \cite{md1}.

We continue with the calculation of the quantum potential for the wave
functions which were found above.

For the wave function $\Psi_{A}\left(  u,v\right)  $ in \cite{isz4} it was
found that there is no nonzero corresponding quantum potential. Wave functions
$\Psi_{B}\left(  u,v\right)  ,~\Psi_{C}\left(  u,v\right)  $ are already
written in polar form with constant amplitudes. Hence the quantum potential
related with these wave functions is zero.

However, from the wave function $\Psi_{D}\left(  u,v\right)  $ and in the
limit $\sqrt{\frac{2}{u}}v\rightarrow\infty$ which is expressed in polar form
from the amplitude $\Omega\left(  u,v\right)  =v^{\frac{\gamma-1}{2}}%
u^{\frac{\gamma+1}{4}}$ the quantum potential term is%
\begin{equation}
Q_{V}\left(  u,v\right)  =\frac{Q_{0}}{uv}~\text{with }Q_{0}=-\frac{\left(
\beta^{2}-1\right)  }{8}\text{. }\label{sz.39}%
\end{equation}
Therefore, for the trajectories with $h=0$, a quantum correction term exists.
Below, we continue our analysis with the study of the effects of the quantum
corrections on the original variables and how the Szekeres system is modified.

\section{The modified Szekeres system}

\label{sec5}

With the use of the quantum potential the Hamiltonian equivalent of the
Szekeres system (\ref{sz.14}) is modified to be%
\begin{equation}
\mathcal{H}=p_{u}p_{v}+\frac{v}{u^{2}}+\frac{Q_{0}}{uv}\equiv0.\label{sz.40}%
\end{equation}
Thus the equations of motion are%
\begin{equation}
\dot{u}=p_{v}~,~\dot{v}=p_{u}~\label{sz.41}%
\end{equation}
and
\begin{equation}
\dot{p}_{v}=-\frac{1}{u^{2}}+\frac{Q_{0}}{uv^{2}}~,~\dot{p}_{v}=2\frac
{v}{u^{3}}+\frac{Q_{0}}{u^{2}v}.\label{sz.42}%
\end{equation}
Therefore with the use of the inverse transformation from the variables
$\left\{  u,v,\dot{u},\dot{v}\right\}  \rightarrow\left\{  \rho,E,\theta
,\sigma\right\}  $ as is given by expressions (\ref{sz.10}), (\ref{sz.10a})
and (\ref{sz.10b}) we find that the only equation of the Szekeres system which
is modified is equation (\ref{sz.02}) which becomes
\begin{equation}
\dot{\theta}+\frac{\theta^{2}}{3}+6\sigma^{2}+\frac{1}{2}\rho+\frac{Q_{0}}%
{72}\frac{\rho^{2}}{E^{2}}\left(  \rho+6E\right)  ^{\frac{4}{3}}%
=0.\label{sz.43}%
\end{equation}

Because we are working in the limit where $\sqrt{\frac{2}{u}}v\rightarrow
\infty$, it follows that $\sqrt{\frac{2}{u}}v=-\frac{6\sqrt{2}E}{\rho}\left(
\frac{\rho}{6}+E\right)  ^{-\frac{1}{6}}\rightarrow\infty$.\ Hence the last
term of (\ref{sz.43}) is approximated as $\frac{Q_{0}}{72}\frac{\rho^{2}%
}{E^{2}}\left(  \rho+6E\right)  ^{\frac{4}{3}}\rightarrow\varepsilon
\frac{Q_{0}}{2}\left(  \rho+6E\right)  $, where $\varepsilon$ is an
infinitesimal parameter such that $\varepsilon^{-1}\rightarrow\infty$.

Thus, equation (\ref{sz.43}) takes the form
\begin{equation}
\dot{\theta}+\frac{\theta^{2}}{3}+6\sigma^{2}+\frac{1}{2}\rho+\frac
{\varepsilon Q_{0}}{2}\left(  \rho+6E\right)  =0.\label{sz.44}%
\end{equation}
We proceed with the study of the evolution of the modified Szekeres system and
we compare our results with that of the original system.

We use the new dimensionless variables in the $\theta$-normalization%
\begin{equation}
\Omega_{m}=\frac{3\rho}{\theta^{2}}~,~\Sigma=\frac{\sigma}{\theta}%
~,~\alpha=\frac{E}{\theta^{2}}~,~\Omega_{R}=\frac{3}{2}\frac{^{\left(
3\right)  }R}{\theta^{2}}.\label{dd.01}%
\end{equation}
The modified Szekeres system is written as follows
\begin{align}
\Omega_{m}^{\prime}  &  =\frac{1}{3}\Omega_{m}\left(  36\Sigma^{2}+\Omega
_{m}-6\Omega_{\Lambda}-1+\varepsilon Q_{0}\left(  \Omega_{m}+18\alpha\right)
\right)  ,\label{dd.02}\\
\Sigma^{\prime}  &  =-\alpha+\frac{1}{6}\Sigma\left(  6\Sigma\left(
1+6\Sigma\right)  -2+\Omega_{m}-6\Omega_{\Lambda}\right)  +\frac{\varepsilon
}{6}Q_{0}\Sigma\left(  \Omega_{m}+18\alpha\right)  ,\label{dd.03}\\
\alpha^{\prime}  &  =\frac{1}{6}\left(  -\Sigma\Omega_{m}+2\alpha\left(
9\Sigma\left(  4\Sigma-1\right)  +\Omega_{m}-6\Omega_{\Lambda}-1\right)
\right)  +\frac{\varepsilon}{3}Q_{0}\alpha\left(  \Omega_{m}+18\alpha
^{2}\right)  ,\label{dd.04}%
\end{align}
where the algebraic equation (\ref{sz.05}) takes the form%
\begin{equation}
\Omega_{R}=9\Sigma^{2}+\Omega_{m}-1\label{dd.06}%
\end{equation}
in which prime \thinspace$"^{\prime}"$ denotes total derivative with respect
to the new independent variable $d\tau=\theta dt.$ The dynamics for the latter
system with $\varepsilon=0$, has been studied in \cite{silent}. However, from
all the stationary points we are interested in the asymptotic solutions which
satisfy the constraint condition $h=0$. Hence from (\ref{sz.14}) and
(\ref{sz.10}), (\ref{sz.10a}), (\ref{sz.10b}) with the use of the dimensional
variables (\ref{dd.01}) the constraint equation is%
\begin{equation}
18\alpha^{2}-\Sigma\left(  1+3\Sigma\right)  \Omega_{m}+\alpha\left(
2-6\Sigma\left(  1+6\Sigma\right)  +\Omega_{m}\right)  =0.\label{dd.07}%
\end{equation}

With the use of the constraint equation (\ref{dd.07}) the three-dimensional
dynamical system (\ref{dd.02})-(\ref{dd.04}) can be reduced by one dimension,
that is, every stationary point $P$ of the system is defined in the
two-dimensional space of variables $P=\left(  \Sigma\left(  P\right)
,\alpha\left(  P\right)  \right)  $ and describes an exact solution with
expansion rate%
\begin{equation}
\frac{\dot{\theta}}{\theta^{2}}=-\frac{1}{6}\left(  2+36\Sigma^{2}+\Omega
_{m}+\varepsilon Q_{0}\left(  \Omega_{m}+18\alpha\right)  \right)  =-q\left(
\Sigma,\alpha,\Omega_{m}\right)  .\label{dd.08}%
\end{equation}
At every point $P$ the expansion rate is found to be $\theta\left(  t\right)
=\frac{1}{q\left(  t-t_{0}\right)  },~$\ for $q\left(  \Sigma\left(  P\right)
,\alpha\left(  P\right)  ,\Omega_{m}\left(  P\right)  \right)  \neq0$ or
$\theta\left(  t\right)  =const.$ for $q\left(  \Sigma\left(  P\right)
,\alpha\left(  P\right)  ,\Omega_{m}\left(  P\right)  \right)  =0.$

\subsection{Stationary points of the Szekeres system}

We study the dynamics of the classic Szekeres system (\ref{dd.03}%
)-(\ref{dd.07}) without the quantum potential term. \ The stationary points
$P=\left(  \Omega_{m}\left(  P\right)  ,\Sigma\left(  P\right)  ,\alpha\left(
P\right)  \right)  $ of the Szekeres system have already been derived in
\cite{silent}. However, because now we impose the constraint (\ref{dd.07}), we
present the analysis in detail.

From the constraint equation (\ref{dd.07}) we see that $\alpha=a\left(
\Omega_{m},\Sigma\right)  $ as follows%
\begin{align}
\alpha_{+}\left(  \Omega_{m},\Sigma\right)   &  =\frac{1}{36}\left(
36\Sigma^{2}+6\Sigma-2-\Omega_{m}+\sqrt{72\Sigma\left(  1+3\Sigma\right)
\Omega_{m}+\left(  2-6\Sigma\left(  1+6\Sigma\right)  +\Omega_{m}\right)
^{2}}\right) \label{dd.09}\\
\alpha_{-}\left(  \Omega_{m},\Sigma\right)   &  =\frac{1}{36}\left(
36\Sigma^{2}+6\Sigma-2-\Omega_{m}-\sqrt{72\Sigma\left(  1+3\Sigma\right)
\Omega_{m}+\left(  2-6\Sigma\left(  1+6\Sigma\right)  +\Omega_{m}\right)
^{2}}\right) \label{dd.10}%
\end{align}

\subsubsection{Branch $\alpha_{+}$}

With the use of $\alpha=\alpha_{+}\left(  \Omega_{m},\Sigma\right)  $ we
obtain a two-dimensional system which admits the following stationary points
$A=\left(  \Omega_{m}\left(  A\right)  ,\Sigma\left(  A\right)  \right)  $,%
\begin{align}
A_{1}  &  =\left(  1,0\right)  ~,~A_{2}=\left(  0,0\right)  ~,~A_{3}=\left(
0,-\frac{1}{3}\right)  ~,~A_{4}=\left(  0,\frac{1}{3}\right)  ~,~\\
A_{5}  &  =\left(  0,\frac{1}{6}\right)  ~,~A_{6}=\left(  -3,-\frac{1}%
{3}\right)  ~,~A_{7}=\left(  -8,-\frac{1}{2}\right)  \text{. }%
\end{align}
Point~$A_{1}$ describes a spatially flat FLRW (-like) universe where the dust
fluid dominates, point $A_{2}$ describes the asymptotic Milne (-like)
universe. The asymptotic solutions at the points $A_{3},~A_{4}$ describe
Bianchi I (-like) spacetimes and specifically Kasner (-like) universes, while
the spacetime at point~$A_{5}$ is that of Kantowski-Sachs (-like) universe.
Points $A_{6},~A_{7}$ are not physically accepted because $\Omega_{m}\left(
A_{6}\right)  <0$ and $\Omega_{m}\left(  A_{7}\right)  <0$.

In order to infer for the stability of the stationary points we determine the
eigenvalues of the linearized systems. The eigenvalues $e\left(  A\right)
=\left(  e_{1}\left(  A\right)  ,e_{2}\left(  A\right)  \right)  $ are,
$e\left(  A_{1}\right)  =\left(  -\frac{1}{2},\frac{1}{3}\right)
$~,~$e\left(  A_{2}\right)  =\left(  -\frac{1}{3},-\frac{1}{3}\right)
$~,~$e\left(  A_{3}\right)  =\left(  1,1\right)  $, $e\left(  A_{4}\right)
=\left(  \frac{7}{3},1\right)  ~$and $e\left(  A_{5}\right)  =\left(
0,\frac{1}{2}\right)  $.\ Therefore we conclude that $A_{2}$ is a stable
point, $A_{1}$ is a saddle point while points $A_{3}$,~$A_{4}$ and $A_{5}$ are
sources. Hence, the unique attractor on that branch is the Milne (-like) universe.

\subsubsection{Branch $\alpha_{-}$}

On the second branch for which $\alpha=\alpha_{-}\left(  \Omega_{m}%
,\Sigma\right)  $, the stationary points of the field equations are
\begin{equation}
B_{1}=\left(  0,-\frac{1}{3}\right)  ~,~B_{2}=\left(  0,\frac{1}{6}\right)
\text{ and }B_{3}=\left(  -8,-\frac{1}{2}\right)  \text{.}%
\end{equation}
Points $B_{1}$ and $B_{2}$ have the same physical properties as $A_{3}$ and
$A_{5}$ respectively, while $B_{3}$ is not physically accepted. As far as the
stability of the stationary points is concerned, we calculate the eigenvalues
of the linearized system around the stationary points from which we get
$e\left(  B_{1}\right)  =\left(  2,1\right)  $ and $e\left(  B_{2}\right)
=\left(  -\frac{1}{2},0\right)  $, and infer that $B_{1}$ is a source,\ while
for point $B_{2}$ in order to infer for the stability we may apply the centre
manifold theorem. The latter is not necessary to be true because we can always
reduce the three-dimensional system (\ref{dd.02}), (\ref{dd.03})
and\ (\ref{dd.04}) into a two-dimensional system for other variables, for the
set $\left\{  \Omega_{m},\alpha\right\}  $ or $\left\{  \Sigma,\alpha\right\}
$ and we can infer the stability properties there. We performed that analysis
and found that point $B_{2}$ is a saddle point.

\begin{figure}[ptb]
\includegraphics[width=1\textwidth]{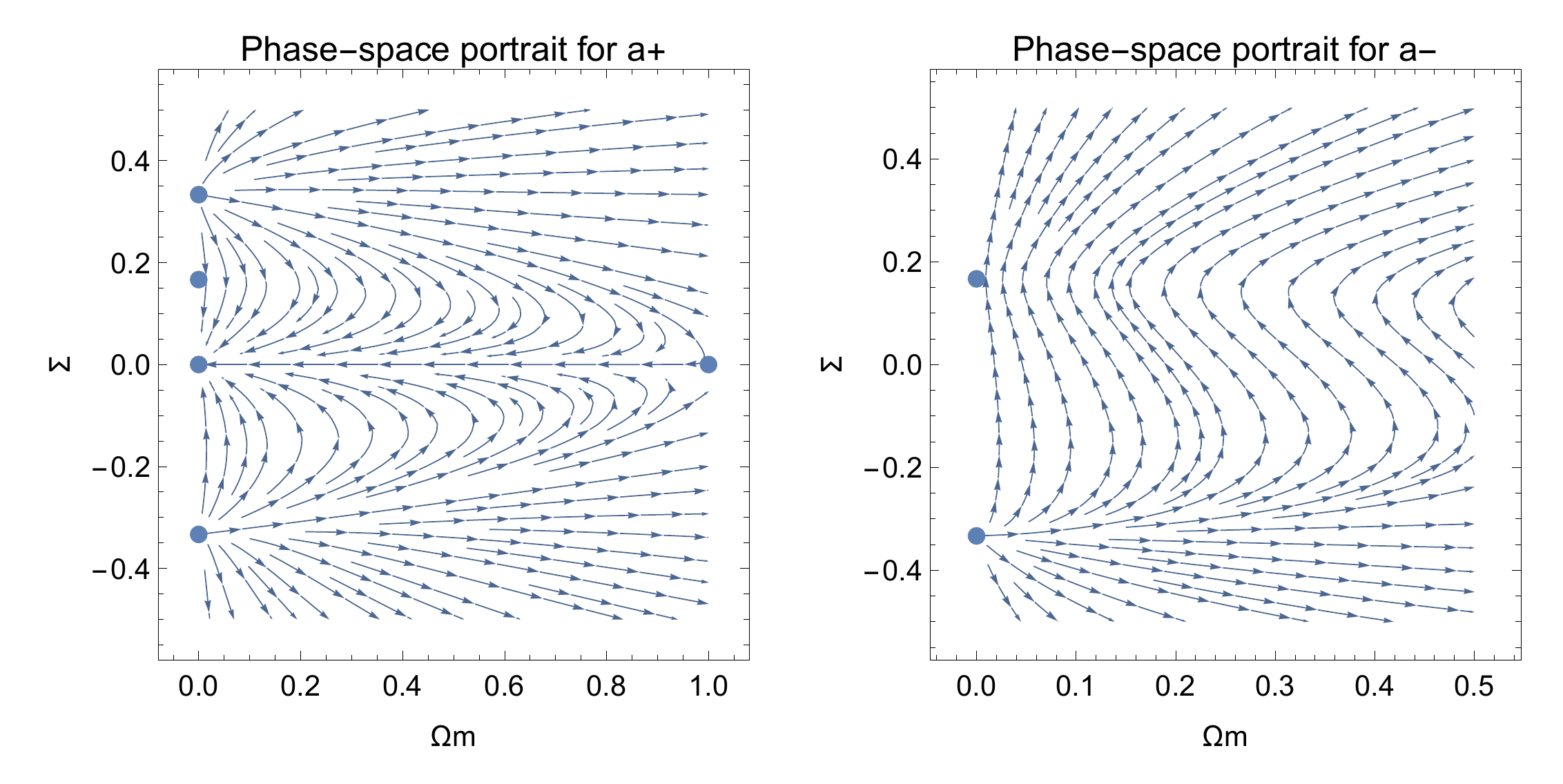}\textbf{\ \newline}%
\caption{Phase-space portraits for the two-dimensional dynamical system
(\ref{dd.02}), (\ref{dd.03}) for $\alpha=\alpha_{+}\left(  \Omega_{m}%
,\Sigma\right)  $ (left figure) and$~\alpha=\alpha_{-}\left(  \Omega
_{m},\Sigma\right)  ~$(right figure). }%
\label{fig2}%
\end{figure}

Therefore the unique attractor for the Szekeres system with the constraint
condition (\ref{dd.07}) is the Milne universe. In Fig. \ref{fig2} we present
the phase-space portrait for the two-dimensional dynamical
system\ (\ref{dd.02}), (\ref{dd.03}) for the two branches $\alpha_{+}$ and
\ $\alpha_{-}$.

\subsection{Stationary points of the modified Szekeres system}

We perform the same analysis for the determination of the stationary points
for the modified Szekeres system with the quantum potential term by assuming
that $\varepsilon Q_{0}\neq0$.

\subsubsection{Branch $\alpha_{+}$}

For the $\alpha_{+}$ branch and for small values of $\varepsilon Q_{0}$, the
stationary points are calculated%
\begin{align}
\bar{A}_{1}  &  =\left(  1-\varepsilon Q_{0},0\right)  ~,~\bar{A}_{2}=\left(
0,0\right)  ~,~\bar{A}_{3}=\left(  0,-\frac{1}{3}\right)  ~,~\bar{A}%
_{4}=\left(  0,\frac{1}{3}\left(  1-\varepsilon Q_{0}\right)  \right)  ~,~\\
\bar{A}_{5}  &  =\left(  0,\frac{1}{6}\right)  ~,~\bar{A}_{6}=\left(
-3,-\frac{1}{3}\right)  ~,~\bar{A}_{7}=\left(  -8+4\varepsilon Q_{0},-\frac
{1}{2}+\frac{\varepsilon Q_{0}}{18}\right)  \text{. }%
\end{align}
Points $\bar{A}_{6},~\bar{A}_{7}$ are not physically accepted, while $\bar
{A}_{1}$ cannot be accepted because at this stationary solution the condition
$\sqrt{\frac{2}{u}}v\rightarrow\infty$ is violated. Therefore, the only
stationary solution which is modified is the asymptotic solution of point
$\bar{A}_{4}$. The spatial curvature at the point $\bar{A}_{4}$ is derived to
be $\Omega_{R}\left(  \bar{A}_{4}\right)  =-\varepsilon Q_{0}$, from which we
infer that the solution at the point describes a Kantowski-Sachs (-like)
universe or a Bianchi III (-like) universe for $\varepsilon Q_{0}<0$. However,
because $\varepsilon Q_{0}\rightarrow0$, the solution at the point can be seen
as the limit of a Kasner (-like) solution

For the asymptotic solution at the points $\bar{A}_{4}$ we calculate $q\left(
\bar{A}_{4}\right)  =-1+\frac{4}{3}\varepsilon Q_{0}$. Hence
\begin{equation}
\frac{\dot{\theta}}{\theta^{2}}=-1+\frac{4}{3}\varepsilon Q_{0}\text{~,~}%
\label{dd.11}%
\end{equation}
that is,%
\begin{equation}
\theta\left(  t\right)  =\frac{1}{1-\frac{4}{3}\varepsilon Q_{0}}\frac
{1}{\left(  t-t_{0}\right)  }\simeq\left(  1+\frac{4}{3}\varepsilon
Q_{0}\right)  \frac{1}{\left(  t-t_{0}\right)  }\label{dd.12}%
\end{equation}
while the anisotropic index $\sigma$ is%
\begin{equation}
\sigma\left(  t\right)  \simeq\frac{1}{3}\left(  1+\frac{\varepsilon Q_{0}}%
{3}\right)  \frac{1}{\left(  t-t_{0}\right)  }.\label{dd.14}%
\end{equation}

Therefore, the power law indices of the scale factors follow from the
algebraic system $p_{1}+2p_{2}=\left(  1+\frac{4}{3}\varepsilon Q_{0}\right)
,~\frac{1}{3}\left(  p_{1}-p_{2}\right)  ^{2}=\frac{1}{3}\left(
1+\frac{\varepsilon Q_{0}}{3}\right)  $which gives%
\begin{equation}
\left(  p_{1},p_{2}\right)  _{A}=\left(  -\frac{1}{3}+\frac{\varepsilon Q_{0}%
}{3},\frac{2}{3}+\frac{\varepsilon Q_{0}}{2}\right)  ~,
\end{equation}%
\begin{equation}
\left(  p_{1},p_{2}\right)  _{B}=\left(  1+\frac{5\varepsilon Q_{0}}{9}%
,\frac{7}{18}\varepsilon Q_{0}\right)  .
\end{equation}

It is clear that the asymptotic solution at point $\bar{A}_{4}$ is not that of
vacuum and that a nonzero energy momentum tensor corresponds to that solution.
In the $1+3$ decomposition, and for the comoving observer $u^{\mu}=\delta
_{t}^{\mu},~\ $the energy momentum tensor has components
\begin{equation}
T_{\mu\nu}=\mu u_{\mu}u_{\nu}+ph_{\mu\nu}+\pi_{\mu\nu},
\end{equation}
where $\mu=T_{\mu\nu}u^{\mu}u^{\nu}~,~p=\frac{1}{3}T_{\mu\nu}h^{\mu\nu}$ and
$\pi_{\mu\nu}=T_{\kappa\lambda}\left(  h_{\mu}^{\kappa}h_{\nu}^{\lambda}%
-\frac{1}{3}h^{\kappa\lambda}h_{\mu\nu}\right)  $.

Hence, for the power indices~$\left(  p_{1},p_{2}\right)  _{A}$ we calculate
\begin{equation}
\mu_{A}=\frac{7}{9}\frac{\varepsilon Q_{0}}{\left(  t-t_{0}\right)  ^{2}%
}~,~p_{A}=-\frac{1}{9}\frac{\varepsilon Q_{0}}{\left(  t-t_{0}\right)  ^{2}}~,
\end{equation}%
\begin{equation}
_{A}\pi_{\nu}^{\mu}=diag\left(  0,-\frac{26}{27}\frac{\varepsilon Q_{0}%
}{\left(  t-t_{0}\right)  ^{2}},\frac{10}{27}\frac{\varepsilon Q_{0}}{\left(
t-t_{0}\right)  ^{2}},\frac{10}{27}\frac{\varepsilon Q_{0}}{\left(
t-t_{0}\right)  ^{2}}\right)  .
\end{equation}

Furthermore, for the set of the power indices $\left(  p_{1},p_{2}\right)
_{B} $ we calculate%
\begin{equation}
\mu_{B}=\frac{7}{9}\frac{\varepsilon Q_{0}}{\left(  t-t_{0}\right)  ^{2}%
}~,~p_{B}=-\frac{1}{9}\frac{\varepsilon Q_{0}}{\left(  t-t_{0}\right)  ^{2}}~,
\end{equation}%
\begin{equation}
_{B}\pi_{\nu}^{\mu}=diag\left(  0,\frac{22}{27}\frac{\varepsilon Q_{0}%
}{\left(  t-t_{0}\right)  ^{2}},-\frac{14}{27}\frac{\varepsilon Q_{0}}{\left(
t-t_{0}\right)  ^{2}},\frac{14}{27}\frac{\varepsilon Q_{0}}{\left(
t-t_{0}\right)  ^{2}}\right)  .
\end{equation}
Recall that in the later solutions $\dot{t}_{0}=0$ holds, but $t_{0}$ is a
function of the space variables of the spacetime.

We conclude that a nonzero energy momentum tensor is introduced by quantum
corrections in Bohmian mechanics for the Szekeres system. Hence ,the Szekeres
system admits semi-classical trajectories which do not remain silent in the
quantum level.

As far as the stability of the stationary points is concerned, that remains
invariant and it is the same with that of the Szekeres system studied above.

\subsubsection{Branch $\alpha_{-}$}

For $\alpha=\alpha_{-}\left(  \Omega_{m},\Sigma\right)  $ we calculate the
stationary points%
\begin{align}
\bar{B}_{1}  &  =\left(  1-Q_{0},0\right)  ~~,~\bar{B}_{2}=\left(  0,-\frac
{1}{3}\right)  ~,~\bar{B}_{3}=\left(  0,\frac{1}{6}\right)  ~,~\\
\bar{B}_{4}  &  =\left(  0,\frac{1}{3}\left(  1-\varepsilon Q_{0}\right)
\right)  ~,~\bar{B}_{5}=\left(  -8+4\varepsilon Q_{0},-\frac{1}{2}%
+\frac{\varepsilon Q_{0}}{18}\right)  .
\end{align}

The stationary point $\bar{B}_{5}$ is not physically accepted, while point
$\bar{B}_{1}$ violates the condition $\sqrt{\frac{2}{u}}v\rightarrow\infty$
and we do not consider it. Points $\bar{B}_{2}~$and $\bar{B}_{3}$ have the
same physical properties and stability with points $B_{2},~B_{3}$
respectively. $\bar{B}_{4}$ is a new point which has the physical properties
of point $\bar{A}_{4}$.

Therefore in the branch $\alpha=\alpha_{-}\left(  \Omega_{m},\Sigma\right)  $
because of the quantum potential a new stationary point exists. The
eigenvalues of the linearized system around the point $\bar{A}_{4}$ are found
to be $e\left(  \bar{A}_{4}\right)  =\left(  1-\frac{4}{3}\varepsilon
Q_{0},\frac{2}{3}-Q_{0}\varepsilon\right)  $ from which we infer that point
$\bar{A}_{4}$ is a source and the asymptotic solution at the point is always unstable.

\section{Conclusion}

\label{sec6}

In this work we studied the quantization process for the Szekeres system and
the effects of the quantum corrections in inhomogeneous cosmological models as
they are described by the De Broglie--Bohm theory for quantum mechanics. We
make use of previous results and we wrote the Szekeres system in its
Hamiltonian equivalent. We calculated the conservation laws for the classical
system and we derived previous results while we were able to determine new
conservation laws which were not found before. One of these conservation laws
is the Lewis invariant which is an important adiabatic invariant for the
oscillator with many applications in quantum mechanics. The remaining new
conservation laws which we derived exist when the Hamiltonian system is
conformally invariant, that is, when the conservation law of the
\textquotedblleft energy\textquotedblright\ is identical to zero. That means
that these new conservation laws, which are constructed by conformal
symmetries of the kinetic metric for the Hamiltonian system, exist for a
specific set of trajectories for the classical Szekeres system.

These new conservation laws applied to define differential operators are
necessary to quantize the Szekeres system. We found new similarity solutions
for the wave function of the Szekeres system. From these new wave functions we
were able to constructed a nonzero quantum potential by applying the approach
of Bohmian mechanics. In order to understand the effects of the quantum
potential term in the original Szekeres system we wrote the modified Szekeres
system and we studied the dynamics and the asymptotic solutions, for the
trajectories with the initial condition the \textquotedblleft
energy\textquotedblright\ of the Hamiltonian equivalent system to be zero.

The main result of this analysis is that because of the quantum correction
term, the dynamical evolution of the Szekeres system is modified, such that
new asymptotic solutions exist which describe approximately Bianchi I
spacetimes which modify the Kasner solutions of the Szekeres systems. These
new Bianchi I solutions correspond to exact inhomogeneous solutions with a
nonzero anisotropic fluid component with nonzero pressure and stress tensor
terms. We conclude that quantum corrections can remove the \textquotedblleft
silent\textquotedblright\ property of the Szekeres universe, making the
Szekeres model more useful in cosmological studies. Therefore, we show that
quantum corrections can be seen as a mechanism to introduce negative pressure
term in the cosmological fluid which is interesting since it may have
applications for the description of the mechanism which starts the
inflationary epoch.

\end{document}